\documentclass[12pt]{article}
\usepackage{amssymb}

\usepackage{graphicx}

\ifx\pdfoutput\undefined
    \relax
\else
    \usepackage{epstopdf}
\fi

\makeatletter
\def\numberbysection{\@addtoreset{equation}{section}
 	\def\theequation{\thesection.\arabic{equation}}}
\makeatother

\numberbysection

\newcommand{\be}{\begin{eqnarray}}
\newcommand{\ee}{\end{eqnarray}}
\newcommand{\non}{\nonumber}

\newcommand{\ch}{\mathop{\rm cosh}\nolimits}
\newcommand{\sh}{\mathop{\rm sinh}\nolimits}

\newcommand{\csch}{\mathop{\rm cosech}\nolimits}

\newcommand{\h}{\ensuremath{\mathsf{h}}}
\newcommand{\sgn}{\mathop{\rm sgn}\nolimits}

\newcommand{\hh}{\mathop{\mathcal H}\nolimits}

\begin{document}

\begin{titlepage}
\strut\hfill UMTG--256
\vspace{.5in}
\begin{center}

\LARGE Boundary {\it S} matrix of an open XXZ spin chain with nondiagonal
boundary terms\\[1.0in]
\large Rajan Murgan\\[0.8in]
\large Physics Department, P.O. Box 248046, University of Miami\\[0.2in]  
\large Coral Gables, FL 33124 USA\\

\end{center}

\vspace{.5in}

\begin{abstract}
Using a recently proposed solution for an open antiferromagnetic spin-$1/2$ XXZ quantum spin chain with $N$ (even) spins and two arbitrary boundary parameters at roots of unity, we compute the boundary scattering amplitudes for one-hole states. We also deduce the relations between the lattice boundary parameters appearing in the spin-chain Hamiltonian and the IR (infrared) parameters that appear in the boundary sine-Gordon {\it S} matrix. 
     
\end{abstract}
\end{titlepage}

\setcounter{footnote}{0}

\section{Introduction}\label{sec:intro}

Factorizable {\it S} matrix is an important object of integrable field theories and integrable quantum spin chains. As for the ``bulk'' case where the {\it S} matrix is determined in terms of two-particle scattering amplitudes, the ``boundary'' case can equally well be formulated in terms of an analogous ``one-particle boundary-reflection'' amplitude. These bulk and boundary amplitudes are required to satisfy Yang-Baxter \cite{Yang}--\cite{Baxter} and the boundary Yang-Baxter \cite{Cherednik,GZ} equations respectively. Methods based on Bethe equations have long been used to compute  bulk two-particle S matrices \cite{Korepin}-- \cite{FT}. In \cite{FT}, Fadeev and Takhtajan studied scattering of spinons for the periodic XXX chain for both the ferromagnetic and antiferromagnetic cases. The bulk two-particle {\it S} matrix for the latter case coincides with the bulk {\it S} matrix for the sine-Gordon model \cite{ZZ} in the limit $\beta^{2}\rightarrow 8\pi$, where $\beta$ is the sine-Gordon coupling constant. Much work has also been done on the subject for open spin chains \cite{S-F}--\cite{doikou} as well as for integrable field theories with boundary \cite{GZ,S-F}. In \cite{GZ}, Ghoshal and Zamolodchikov presented a precise formulation of the concept of boundary {\it S} matrix for 1 + 1 dimensional quantum field theory with boundaries such as Ising field theory with boundary magnetic field and boundary sine-Gordon model. For the latter model, the authors used a bootstrap approach to compute the boundary {\it S} matrix. They determined the scalar factor up to a CDD-type of ambiguity. Nonlinear integral equation (NLIE) \cite{KB, Destri} approach has also been used to study excitations in integrable quantum field theories such as the sine-Gordon model \cite{Fio}--\cite{Feverati} and open quantum spin-$1/2$ XXZ spin chains \cite{LMSS}--\cite{ABNPT}. In fact, in \cite{ABNPT}, NLIE approach is used to compute boundary {\it S} matrix for the open spin-$1/2$ XXZ spin chain with nondiagonal boundary terms, where the boundary parameters obey certain constraint. The bulk anisoptopy parameter however is taken to be arbitrary. 

In this paper, we compute the eigenvalues of the boundary {\it S} matrix for a special case of an open spin-$1/2$ XXZ spin chain with nondiagonal boundary terms with two independent boundary parameters (with no constraint) at roots of unity, using the solution obtained recently \cite{MN,MNS}. The motivation for the performed computation is the fact that the Bethe Ansatz equation for this model is unchanged under sign reversal of the boundary parameters. Hence, the usual trick of obtaining the second eigenvalue of the boundary $S$ matrix of an open spin-$1/2$ XXZ spin chain by exploiting the change in Bethe Ansatz equation under such sign reversal of the boundary parameters \cite{GMN, AN, ABNPT, doikou} would not work here. Consequently, identifications of separate one-hole states are necessary here. As far as the formalism goes, we follow the approach used earlier for diagonal open spin chains \cite{GMN, AN}. This is a generalization of the method developed by Korepin, Andrei and Destri \cite{Korepin,A-D} for computing bulk {\it S} matrix. The quantization condition discussed by Fendley and Saleur \cite{S-F} is a crucial step for the calculation. The solution utilized here was derived for certain values of bulk anisotropy parameter, $\mu$ in the repulsive regime $(\mu = {\pi\over p+1}\in (0,{\pi\over 2}])$ for odd $p$ values. Hence, we focus only on the critical and repulsive regime, which corresponds in the sine-Gordon model to $\beta^{2}\in[4\pi, 8\pi)$ \footnote{$\beta^{2} = 8(\pi-\mu)$}. One-hole excitations for this model occur in even {\it N} sector \cite{Murgan} in contrast to the diagonal open spin-$1/2$ XXZ spin chain where such excitations appear in the odd {\it N} sector \cite{AN}.  

The outline of the paper is as follows. In Section 2, we briefly review the model. Previously found Bethe Ansatz solution for the model is presented here \cite{MN,MNS}. We also review the string hypothesis for two one-hole states. In Section 3, we proceed with the computation of the scattering amplitudes. Since the Bethe roots for the model consist of ``sea'' roots and ``extra'' roots, we rely on a conjectured relation between the ``extra'' roots and the hole rapidity, which is confirmed numerically for system up to about 60 sites. We find that the eigenvalue derived for the open XXZ spin chain agrees with one of the eigenvalues of Ghoshal-Zamolodchikov's boundary {\it S} matrix for the one boundary sine-Gordon model, provided the lattice boundary parameters that appear in the spin chain Hamiltonian and the IR parameters that appear in Ghoshal-Zamolodchikov's boundary {\it S} matrix \cite{GZ} obey the same relation as in \cite{ABNPT} \footnote{Very recently, similar relations were found for the open XXZ spin chain with diagonal-nondiagonal boundary terms in \cite{doikou}.}. The problem of finding the second eigenvalue of the boundary {\it S} matrix requires the identification of an independent one-hole state. In contrast to previous studies \cite{GMN, AN, ABNPT, doikou}, where such state was found by reversing the signs of the boundary parameters \footnote{In fact, there is a change $\xi_{\pm}\rightarrow -\xi_{\pm}$ in the Bethe equation for the diagonal case \cite{AN}.}, similar strategy does not work here. Reversing the signs of the boundary parameters in the present case leaves the Bethe equation unchanged, hence giving the same one-hole state. Interestingly, a separate one-hole state with 2-string is found \cite{Murgan}. Using a conjectured relation between ``extra'' roots, hole rapidity and the boundary parameters, which is again confirmed numerically for system up to about 60 sites, we derive the remaining eigenvalue which also agrees with Ghoshal-Zamolodchikov's result. Finally, we conclude the paper with a brief discussion and possible future work on the subject in Section 4.                   

\section{Bethe Ansatz and string hypothesis}\label{sec:Bethe}

We begin this section by reviewing recently proposed Bethe Ansatz solution \cite{MN,MNS} for 
the following model \cite{GZ,dVega}
\be
\hh &=& \hh_{0}
+ {1\over 2}\sh \eta \big( 
 \csch \alpha_{-}\sigma_{1}^{x} 
 + \csch \alpha_{+}\sigma_{N}^{x}
\big)  \,, \label{Hamiltonian} 
\ee
where the ``bulk'' Hamiltonian is given by
\be 
\hh_{0} = {1\over 2}\sum_{n=1}^{N-1}\left( 
\sigma_{n}^{x}\sigma_{n+1}^{x}+\sigma_{n}^{y}\sigma_{n+1}^{y}
+\ch \eta\ \sigma_{n}^{z}\sigma_{n+1}^{z}\right) \,. 
\label{bulkHamiltonian}
\ee
In the above expressions, $\sigma^{x}$, $\sigma^{y}$, $\sigma^{z}$ are the usual
Pauli matrices, $\eta$ is the bulk anisotropy parameter (taking values $\eta = {i\pi\over{p+1}}$, with $p$ odd),
$\alpha_{\pm}$ are the boundary parameters, and $N$ is the number of spins/sites. 
Note that this model has only two boundary parameters. The most general integrable boundary terms contain six boundary parameters. In the present case, four other boundary parameters
have been set to zero. We restrict the values of the remaining parameters, $\alpha_{\pm}$ to be pure imaginary 
to ensure the hermiticity of the Hamiltonian (\ref{Hamiltonian}). The Bethe Ansatz equations are given by 
\be
{\delta(u_{j}^{(1)})\ h^{(2)}(u_{j}^{(1)}-\eta)
\over \delta(u_{j}^{(1)}-\eta)\ h^{(1)}(u_{j}^{(1)})} 
&=&-{Q_{2}(u_{j}^{(1)}-\eta)\over Q_{2}(u_{j}^{(1)}+\eta)} \,, \qquad j =
1\,, 2\,, \ldots \,, M_{1} \,, \non \\
{h^{(1)}(u_{j}^{(2)}-\eta)\over h^{(2)}(u_{j}^{(2)})}
&=&-{Q_{1}(u_{j}^{(2)}+\eta)\over Q_{1}(u_{j}^{(2)}-\eta)} \,, \qquad j =
1\,, 2\,, \ldots \,, M_{2} \,.
\label{BAEII1}
\ee
where 
\be
\delta(u) &=& 2^{4}\left( \sinh u \sinh(u + 2\eta) \right)^{2N} {\sinh 2u
\sinh (2u + 4\eta)\over \sinh(2u+\eta) \sinh(2u+3\eta)}
\sinh(u+\eta+\alpha_{-})\non \\
& & \sinh(u+\eta-\alpha_{-})\sinh(u+\eta+\alpha_{+})\sinh(u+\eta-\alpha_{+})\cosh^{4}(u + \eta)
\label{delta}\,,
\ee 
\be
h^{(1)}(u) &=& {8\sinh^{2N+1}(u+2\eta)\cosh^{2}(u+\eta) \cosh(u+2\eta)\over 
\sinh(2u+3\eta)} \,, \quad h^{(2)}(u) = h^{(1)}(-u-2\eta) \label{h1function}
\ee
and 
\be
Q_{a}(u) = \prod_{j=1}^{M_{a}} 
\sinh (u - u_{j}^{(a)}) \sinh (u + u_{j}^{(a)} + \eta) \,, \qquad a = 
1\,, 2\,, 
\label{QII}
\ee
$M_{1}$ and $M_{2}$ are the number of Bethe roots, $u_{j}^{(1)}$ and $u_{j}^{(2)}$ 
(zeros of $Q_{1}(u)$ and $Q_{2}(u)$ respectively). 

\subsection{One-hole state}\label{sec:one-hole}

In order to compute the spinon boundary scattering amplitude, we consider a one-hole state. The roots distribution for such a state was found in \cite{MNS}. One-hole excitations for the open XXZ spin chain we study here appear in the even $N$ sector. Hence, it is sufficient to review the results for even $N$ case. The shifted Bethe roots $\tilde u_{j}^{(a)} = u_{j}^{(a)} + {\eta\over 2}$ for this state have the following form
\be
\left\{ \begin{array}{c@{\quad : \quad} l}
\mu \lambda_{j}^{(a,1)}                      & j = 1\,, 2\,, \ldots \,, M_{(a,1)} \\
\mu \lambda_{j}^{(a,2)} + {i \pi\over 2} \,, & j = 1\,, 2\,, \ldots \,, M_{(a,2)}
\end{array} \right. \,, \qquad a = 1\,, 2 \,, 
\label{stringhypothesisIIc}
\ee 
where $\mu = {\pi\over{p+1}}$ and $\lambda_{j}^{(a,b)}$ are real. Here, $M_{(1,1)}= M_{(2,1)} = {N\over 2}$, and $M_{(1,2)}={p+1\over 2}$, $M_{(2,2)}={p-1\over 2}$.
The $\mu\lambda_{j}^{(a,1)}$ are the zeros of $Q_{a}(u)$ that form real sea (``sea'' roots) and $\mu\lambda_{k}^{(a,2)}$ are 
real parts of the ``extra'' roots (also zeros of $Q_{a}(u)$) which are not part of the ``seas''. 
Hence, there are two ``seas'' of real roots. We employ notations used in \cite{Anep}, 
\be
e_{n}(\lambda) =
{\sinh \left(\mu  (\lambda + {i n\over 2}) \right) 
\over \sinh \left( \mu (\lambda - {i n\over 2}) \right) } \,, \qquad
g_{n}(\lambda) = e_{n}(\lambda \pm {i \pi \over 2 \mu})
= {\cosh \left(\mu  (\lambda + {i n\over 2}) \right) 
\over \cosh \left( \mu (\lambda - {i n\over 2}) \right) } \,.
\label{eandgfunctions}
\ee
Rewriting bulk and boundary parameters \cite{Anep}, $\eta = i\mu$, 
$\alpha_{\pm} = i\mu a_{\pm}$\footnote{The string hypothesis (\ref{stringhypothesisIIc}) holds true only for suitable values of $a_{\pm}$, namely ${\nu-1 \over 2} < |a_{\pm}| < {\nu+1 \over 2}$ \,, 
\quad $a_{+} a_{-} > 0$\,, where $\nu = p+1$} 
the Bethe Ansatz equations (\ref{BAEII1}) for the ``sea'' roots then take the following form
\be
\lefteqn{e_{1}(\lambda_{j}^{(1,1)})^{2N+1} 
\left[ g_{1}(\lambda_{j}^{(1,1)}) 
e_{1+2a_{-}}(\lambda_{j}^{(1,1)}) e_{1-2a_{-}}(\lambda_{j}^{(1,1)})
e_{1+2a_{+}}(\lambda_{j}^{(1,1)}) e_{1-2a_{+}}(\lambda_{j}^{(1,1)}) \right]^{-1}} 
\label{BAEIIsea1} \\
& & = -\prod_{k=1}^{N/2} \left[
e_{2}(\lambda_{j}^{(1,1)} - \lambda_{k}^{(2,1)}) 
e_{2}(\lambda_{j}^{(1,1)} + \lambda_{k}^{(2,1)})\right] 
 \prod_{k=1}^{(p-1)/2} \left[ 
g_{2}(\lambda_{j}^{(1,1)} - \lambda_{k}^{(2,2)}) 
g_{2}(\lambda_{j}^{(1,1)} + \lambda_{k}^{(2,2)}) \right] \,, \non 
\ee 
and
\be
\lefteqn{e_{1}(\lambda_{j}^{(2,1)})^{2N+1} 
g_{1}(\lambda_{j}^{(2,1)})^{-1}} 
\label{BAEIIsea2} \\
& & = -\prod_{k=1}^{N/2} \left[
e_{2}(\lambda_{j}^{(2,1)} - \lambda_{k}^{(1,1)}) 
e_{2}(\lambda_{j}^{(2,1)} + \lambda_{k}^{(1,1)})\right] 
\prod_{k=1}^{(p+1)/2} \left[  
g_{2}(\lambda_{j}^{(2,1)} - \lambda_{k}^{(1,2)}) 
g_{2}(\lambda_{j}^{(2,1)} + \lambda_{k}^{(1,2)}) \right] \,, \non 
\ee 
respectively, where $j=1\,, \ldots \,, {N\over 2}$. These equations can be re-expressed in terms of counting functions, $\h^{(l)}(\lambda)$ as 
\be
\h^{(l)}(\lambda_{j}^{(l,1)}) = J_{j}, \quad l = 1\,,2
\label{BAcounting}
\ee 
where $\h^{(l)}(\lambda)$ are given by
\be
\lefteqn{\h^{(1)}(\lambda) = {1\over 2 \pi}\Big\{ (2N+1) 
q_{1}(\lambda) - r_{1}(\lambda)
- q_{1+2a_{-}}(\lambda) - q_{1-2a_{-}}(\lambda)
- q_{1+2a_{+}}(\lambda) - q_{1-2a_{+}}(\lambda)} \non \\  
& & -\sum_{k=1}^{N/2}\left[ 
q_{2}(\lambda - \lambda_{k}^{(2,1)}) +
q_{2}(\lambda + \lambda_{k}^{(2,1)})\right] 
- \sum_{k=1}^{(p-1)/2} \left[
r_{2}(\lambda - \lambda_{k}^{(2,2)}) +
r_{2}(\lambda + \lambda_{k}^{(2,2)}) \right] \Big\} \,,
\label{h1even}
\ee 
and
\be
\lefteqn{\h^{(2)}(\lambda) = {1\over 2 \pi}\Big\{ (2N+1) 
q_{1}(\lambda) - r_{1}(\lambda)} \non \\
& & -\sum_{k=1}^{N/2}\left[ 
q_{2}(\lambda - \lambda_{k}^{(1,1)}) +
q_{2}(\lambda + \lambda_{k}^{(1,1)})\right] 
- \sum_{k=1}^{(p+1)/2} \left[
r_{2}(\lambda - \lambda_{k}^{(1,2)}) +
r_{2}(\lambda + \lambda_{k}^{(1,2)}) \right] \Big\} \,.
\label{h2even}
\ee 
In the above equations, $q_{n}(\lambda)$ and $r_{n}(\lambda)$ are odd functions defined
by
\be
q_{n}(\lambda) &=& \pi + i \ln e_{n}(\lambda) 
= 2 \tan^{-1}\left( \cot(n \mu/ 2) \tanh( \mu \lambda) \right)
\,, \non \\
r_{n}(\lambda) &=&  i \ln g_{n}(\lambda) \,.
\label{logfuncts}
\ee
Further, $\big\{J_{1}, J_{2},\ldots, J_{{N\over 2}}\big\}$ is a set of increasing positive integers that parametrize the state \footnote{In principle, there are two such sets of integers, $\big\{J_{i}^{(1)}\big\}$ and $\big\{J_{i}^{(2)}\big\}$ corresponding to the two counting functions, $\h^{(1)}(\lambda)$ and $\h^{(2)}(\lambda)$ respectively. But, in fact these two sets of integers are identical. Hence we choose to drop the superscript, $l$ from $J_{j}$ in (\ref{BAcounting}).}. For states with no holes, the integers take consecutive values. For one-hole state, there is a break in the sequence, represented by a missing integer. This missing integer $\tilde{J}$, fixes the value of the hole rapidity, $\tilde{\lambda}$, according to
\be
\h^{(1)}(\tilde{\lambda}) = \h^{(2)}(\tilde{\lambda}) = \tilde{J}.
\label{hole}
\ee
If the hole is located to the right of the largest ``sea'' root ($\lambda_{{N\over 2}}^{(a,1)}$), then $\tilde{J} = \lfloor\h^{(l)}(\infty) - \h^{(l)}(\lambda_{{N\over 2}}^{(a,1)})\rfloor$. See \cite{Murgan} for more details. For later use, we next define the densities of ``sea'' roots as 
\be
\rho^{(l)}(\lambda) = {1\over N}{d\h^{(l)}(\lambda)\over d\lambda}
\label{dens}
\ee
where
$l = 1\,,2$
 
\noindent The functions (\ref{logfuncts}) have the following derivatives which prove to be essential to the analysis in following sections,
\be
a_n(\lambda) &=& {1\over 2\pi} {d \over d\lambda} q_n (\lambda)
= {\mu \over \pi} 
{\sin (n \mu)\over \cosh(2 \mu \lambda) - \cos (n \mu)} \,, \non \\
b_n(\lambda) &=& {1\over 2\pi} {d \over d\lambda} r_n (\lambda)
= -{\mu \over \pi} 
{\sin (n \mu)\over \cosh(2 \mu \lambda) + \cos (n \mu)} \,. 
\label{anbn}
\ee
 
\subsection{One-hole state with 2-string}\label{sec:2string}

In addition to the one-hole state considered in last section, there is another one-hole state. This state is the only remaining one-hole state, which also has a 2-string. In this section, we give some brief information on the state. The shifted Bethe roots $\tilde u_{j}^{(a)} = u_{j}^{(a)} + {\eta\over 2}$ for this state have the following form
\be
\left\{ \begin{array}{c@{\quad  \quad} l}
\mu \lambda_{j}^{(a,1)}                      & j = 1\,, 2\,, \ldots \,, M_{(a,1)} \\
\mu \lambda_{j}^{(a,2)} + {i \pi\over 2} \,, & j = 1\,, 2\,, \ldots \,, M_{(a,2)} \\
\mu \lambda_{0}^{(a)} + {\eta\over 2} \\
\mu \lambda_{0}^{(a)} - {\eta\over 2} \\
\end{array} \right. \,, \qquad a = 1\,, 2 \,, 
\label{stringhypothesisIId}
\ee
where $\lambda_{0}^{(a)}$\,, $\mu = {\pi\over{p+1}}$ and $\lambda_{j}^{(a,b)}$ are real. Here, $M_{(1,1)}= M_{(2,1)} = {N\over 2}-1$, and $M_{(1,2)}={p-1\over 2}$, $M_{(2,2)}={p-3\over 2}$. As before, $\mu\lambda_{j}^{(a,1)}$ are the zeros of $Q_{a}(u)$ that form real sea (``sea'' roots) and $\mu\lambda_{k}^{(a,2)}$ are 
real parts of the ``extra'' roots (also zeros of $Q_{a}(u)$) which are not part of the ``seas''. For this state, we also have $\mu\lambda_{0}^{(a)}$, the real parts of additional ``extra'' roots that form a 2-string. 
 
The counting functions for this state are given by 
\be
\lefteqn{\h^{(1)}(\lambda) = {1\over 2 \pi}\Big\{ (2N+1) 
q_{1}(\lambda) - r_{1}(\lambda)
- q_{1+2a_{-}}(\lambda) - q_{1-2a_{-}}(\lambda)
- q_{1+2a_{+}}(\lambda) - q_{1-2a_{+}}(\lambda)} \non \\  
& & -\sum_{k=1}^{{N\over 2} - 1}\left[ 
q_{2}(\lambda - \lambda_{k}^{(2,1)}) +
q_{2}(\lambda + \lambda_{k}^{(2,1)})\right] 
- \sum_{k=1}^{(p-3)/2} \left[
r_{2}(\lambda - \lambda_{k}^{(2,2)}) +
r_{2}(\lambda + \lambda_{k}^{(2,2)}) \right] \non \\
& & - q_{3}(\lambda - \lambda_{0}^{(2)}) - q_{3}(\lambda + \lambda_{0}^{(2)}) - q_{1}(\lambda - \lambda_{0}^{(2)}) - q_{1}(\lambda + \lambda_{0}^{(2)}) \Big\} \,,
\label{h1even2}
\ee 
and
\be
\lefteqn{\h^{(2)}(\lambda) = {1\over 2 \pi}\Big\{ (2N+1) 
q_{1}(\lambda) - r_{1}(\lambda)} \non \\
& & -\sum_{k=1}^{{N\over 2} -1}\left[ 
q_{2}(\lambda - \lambda_{k}^{(1,1)}) +
q_{2}(\lambda + \lambda_{k}^{(1,1)})\right] 
- \sum_{k=1}^{(p-1)/2} \left[
r_{2}(\lambda - \lambda_{k}^{(1,2)}) +
r_{2}(\lambda + \lambda_{k}^{(1,2)}) \right] \non \\
& & - q_{3}(\lambda - \lambda_{0}^{(1)}) - q_{3}(\lambda + \lambda_{0}^{(1)}) - q_{1}(\lambda - \lambda_{0}^{(1)}) - q_{1}(\lambda + \lambda_{0}^{(1)}) \Big\} \,.
\label{h2even2}
\ee
The Bethe Ansatz equations for this state take the following form, 
\be
\h^{(l)}(\lambda_{j}^{(l,1)}) = J_{j},\quad   l = 1\,,2
\label{BAcounting2}
\ee 
where $\big\{J_{1}, J_{2},\ldots, J_{{N\over 2}-1}\big\}$ is a set of increasing positive integers that parametrize the state. The hole for this state breaks the sequence, represented by a missing integer. As before, the missing integer $\tilde{J}$,enables one to calculate the hole rapidity, $\tilde{\lambda}$ using 
\be
\h^{(1)}(\tilde{\lambda}) = \h^{(2)}(\tilde{\lambda}) = \tilde{J}.
\label{hole}
\ee
If the hole appears to the right of the largest ``sea'' root ($\lambda_{{N\over 2}-1}^{(a,1)}$), then $\tilde{J} = \lfloor\h^{(l)}(\infty) - \h^{(l)}(\lambda_{{N\over 2}-1}^{(a,1)})\rfloor$. More on this state can be found in \cite{Murgan}.

\section{Boundary {\it S} matrix}\label{sec:smatrix}

In this Section, we give the derivation for the boundary scattering amplitudes for one-hole states reviewed in Section \ref{sec:Bethe}. 

\subsection{Eigenvalue for the one-hole state without 2-string}\label{wo2s}

First, we consider the state reviewed in Section \ref{sec:one-hole}. From (\ref{h1even}), (\ref{h2even}), (\ref{dens}) and (\ref{anbn}), one can solve for the sum of the two densities. We recall the results below \cite{MNS},
\be
\rho_{total}(\lambda) &=& \rho^{(1)}(\lambda) + \rho^{(2)}(\lambda)\non \\
&=& 4 s(\lambda) + {1\over N} R_{+}(\lambda)           
\label{solutionfordensity}
\ee  
where $s(\lambda) = {1\over 2 \ch(\pi\lambda)}$ and $R_{+}(\lambda)$ is the inverse Fourier transform of $\hat R_{+}(\omega)$ \footnote{Our conventions are
\be
\hat f(\omega) \equiv \int_{-\infty}^\infty e^{i \omega \lambda}\ 
f(\lambda)\ d\lambda \,, \qquad\qquad
f(\lambda) = {1\over 2\pi} \int_{-\infty}^\infty e^{-i \omega \lambda}\ 
\hat f(\omega)\ d\omega \,. \non 
\ee} 
which is given by
\be
\hat R_{+}(\omega) &=& {1\over 1+\hat a_{2}(\omega)}
\Big[2 \hat a_{1}(\omega) + 2 \hat a_{2}(\omega) - 2 \hat b_{1}(\omega)
- \hat a_{1+2a_{-}}(\omega) - \hat a_{1-2a_{-}}(\omega) - \hat a_{1+2a_{+}}(\omega) - \hat a_{1-2a_{+}}(\omega) \non \\ 
&-&  2\hat b_{2}(\omega)\big(\sum_{k=1}^{{p-1\over 2}}\cos(\lambda_{k}^{(2,2)}\omega) + \sum_{l=1}^{{p+1\over 2}}\cos(\lambda_{l}^{(1,2)}\omega)\big)+ 4 \hat a_{2}(\omega)\cos(\tilde{\lambda} \omega) \Big]
\label{totden}
\ee
and
\be
\hat a_{n}(\omega) &=& \sgn(n) {\sinh \left( (\nu  - |n|) 
\omega / 2 \right) \over
\sinh \left( \nu \omega / 2 \right)} \,,
\qquad 0 \le |n| < 2 \nu  \,, \label{fourier1} \\
\hat b_{n}(\omega) &=&
-{\sinh \left( n \omega / 2 \right) \over
\sinh \left( \nu \omega / 2 \right)} \,,
\qquad \qquad\qquad\quad  0 < \Re e\ n < \nu  \,.
\label{fourier2}
\ee
are the Fourier transforms of (\ref{anbn}).
The presence of ``extra'' roots, $\lambda_{k}^{(a,2)}$ and the hole rapidity, $\tilde{\lambda}$, are to be noted here\footnote{Energy carried by the hole is given by $E(\tilde{\lambda}) = {\pi\sin\mu\over 2 \mu}{1\over \cosh(\pi\tilde{\lambda})}$. Such an expression for spinon was derived in \cite{FT}}. Henceforth, we shall denote $\lambda_{k}^{(a,2)}$ simply as $\lambda_{k}^{(a)}$. Morever, momentum of the excitation is given by
\be
p(\tilde{\lambda}) =  \tan^{-1}\left( \sh(\pi \tilde{\lambda}) \right) - {\pi\over 2}           
\label{momentum}
\ee
From (\ref{momentum}), one gets $s(\lambda) = {1\over 2\pi }{dp(\lambda)\over d\lambda}$. Consequently, using (\ref{dens}), one rewrites (\ref{solutionfordensity}) as
\be
{1\over N}{d \h_{total}(\lambda)\over d\lambda} = {2\over \pi}{dp(\lambda)\over d\lambda} + {1\over N} R_{+}(\lambda)
\label{Yang} 
\ee
where $\h_{total}(\lambda)=\h^{(1)}(\lambda)+\h^{(2)}(\lambda)$ and ${1\over N}{d \h_{total}(\lambda)\over d\lambda} = \rho_{total}(\lambda)$.     
After integrating (\ref{Yang}) with respect to $\lambda$, taking limits of integration from $0$ to $\tilde{\lambda}$, one finds \footnote{Since we are only able to determine the scattering amplitudes up to a rapidity-independent factor, the additive constant $p(0)$ from the integration is ignored in (\ref{Yang2}).}
\be
\h_{total}(\tilde{\lambda}) &=& \h^{(1)}(\tilde{\lambda}) + \h^{(2)}(\tilde{\lambda})
= {2\over \pi}Np(\tilde{\lambda}) + \int_{0}^{\tilde{\lambda}} d\lambda R_{+}(\lambda) 
\label{Yang2} 
\ee
Since $\h^{(1)}(\tilde{\lambda}) = \h^{(2)}(\tilde{\lambda})\in$ positive integer and $R_{+}(\lambda)$ is an even function of $\lambda$, multiplying the resulting expression by $2i\pi$ and exponentiating gives 
\be
e^{2ip(\tilde{\lambda})N}e^{{i\pi\over 2}\int_{-\tilde{\lambda}}^{\tilde{\lambda}} d\lambda R_{+}(\lambda)} = 1
\label{Yang3}
\ee
Next, let us compare equation (\ref{Yang3}) to the Yang's quantization condition for a particle on an interval of length $N$,
\be
e^{2ip(\tilde{\lambda})N}R(\tilde{\lambda};a_{+})R(\tilde{\lambda};a_{-})\mid\tilde{\lambda},(\pm)\rangle = \mid\tilde{\lambda},(\pm)\rangle
\label{YangMatrix}
\ee
where $R(\tilde{\lambda};a_{\pm})$ are the non-diagonal boundary {\it S} matrices and $\mid\tilde{\lambda},(\pm)\rangle$ denote the two possible one-hole states. Note that the $\pm$ in $\mid\tilde{\lambda},(\pm)\rangle$ represents two posssible one-hole states and not the right and left boundaries. The expression $e^{{i\pi\over 2}\int_{-\tilde{\lambda}}^{\tilde{\lambda}} d\lambda R_{+}(\lambda)}$ then, should be equal to one of the two eigenvalues of the Yang matrix $Y(\tilde{\lambda})$ defined by
\be
Y(\tilde{\lambda}) = R(\tilde{\lambda};a_{+})R(\tilde{\lambda};a_{-})
\label{YangMatrix2}
\ee
Defining this eigenvalue as 
$\alpha(\tilde{\lambda},a_{+})\alpha(\tilde{\lambda},a_{-})$, where $+$ and $-$ denote the right and left boundaries respectively, (\ref{Yang3}) can be rephrased as
\be
e^{2ip(\tilde{\lambda})N}\alpha(\tilde{\lambda},a_{+})\alpha(\tilde{\lambda},a_{-}) = 1
\label{Yang4}
\ee 
The problem thus reduces to evaluating the following
\be
\alpha(\tilde{\lambda},a_{+})\alpha(\tilde{\lambda},a_{-}) = e^{{i\pi\over 2}\int_{-\tilde{\lambda}}^{\tilde{\lambda}} d\lambda R_{+}(\lambda)}
\label{amplitudes}
\ee
After some manipulations, we have the following,
\be
\alpha(\tilde{\lambda},a_{+})\alpha(\tilde{\lambda},a_{-}) &=& \exp\bigg\{2\int_{0}^{\infty}{d\omega\over \omega}\sh(2i\tilde{\lambda}\omega)\Big[{\hat a_{2}(\omega)\over 1+\hat a_{2}(\omega)} + {1\over 1+\hat a_{2}(2\omega)}\big[\hat a_{2}(2\omega) + \hat a_{1}(2\omega) - \hat b_{1}(2\omega)\non \\ 
&-& {1\over 2}(\hat a_{1+2a_{-}}(2\omega) + \hat a_{1-2a_{-}}(2\omega) + \hat a_{1+2a_{+}}(2\omega) + \hat a_{1-2a_{+}}(2\omega))\non \\ 
&-& \hat b_{2}(2\omega)(\sum_{k=1}^{{p-1\over 2}}\cos(2\lambda_{k}^{(2)}\omega) + \sum_{l=1}^{{p+1\over 2}}\cos(2\lambda_{l}^{(1)}\omega))\big]\Big]\bigg\}
\label{amplitudes2}
\ee
Further, using (\ref{fourier1}) and (\ref{fourier2}), one gets
\be
\alpha(\tilde{\lambda},a_{+})\alpha(\tilde{\lambda},a_{-}) &=& \exp\bigg\{2\int_{0}^{\infty}{d\omega\over \omega}\sh(2i\tilde{\lambda}\omega)\Big[{2\sh(3\omega/2)\sh((\nu-2)\omega/2)\over \sh(2\omega)\sh((\nu-1)\omega/2)}\non \\ 
&+& {\sh(\omega)\over \sh((\nu-1)\omega)\ch(\omega)} + {\sh((-\nu+2a_{-}-1)\omega)\over 2\sh((\nu-1)\omega)\ch(\omega)}\non \\ &+& {\sh((\nu-2a_{-}-1)\omega)\over 2\sh((\nu-1)\omega)\ch(\omega)} + (a_{-}\rightarrow a_{+})\non \\ 
&+& {\sh(\omega)\over \sh((\nu-1)\omega)}(\sum_{k=1}^{{p-1\over 2}}\cos(2\lambda_{k}^{(2)}\omega) + \sum_{l=1}^{{p+1\over 2}}\cos(2\lambda_{l}^{(1)}\omega))\Big]\bigg\}
\label{amplitudes3}
\ee
where $(a_{-}\rightarrow a_{+})$ is a shorthand for two additional terms which are the same as the third and fourth terms in the integrand of (\ref{amplitudes3}), but with $a_{-}$ replaced by $a_{+}$.  
The integrals involving ``extra'' roots $\lambda_{k}^{(2)}$ and $\lambda_{l}^{(1)}$ yield
\be
\exp\bigg\{2\int_{0}^{\infty}{d\omega\over \omega}\sh(2i\tilde{\lambda}\omega){\sh(\omega)\over \sh((\nu-1)\omega)}(\sum_{k=1}^{{p-1\over 2}}\cos(2\lambda_{k}^{(2)}\omega) + \sum_{l=1}^{{p+1\over 2}}\cos(2\lambda_{l}^{(1)}\omega))\bigg\} \non \\
= \prod_{k=1}^{{p-1\over 2}}\prod_{l=1}^{{p+1\over 2}}\sqrt{f(\lambda_{k}^{(2)},\lambda_{l}^{(1)},\tilde{\lambda})f(\lambda_{k}^{(2)},\lambda_{l}^{(1)},\tilde{\lambda}+{i\pi\over \mu'})}
\label{constraint}
\ee
where $\mu' = {\pi\over \nu-1}$ and
\be
f(\lambda_{k}^{(2)},\lambda_{l}^{(1)},\tilde{\lambda}) &=& {\sh\big({\mu'\over 2}(\tilde{\lambda}+\lambda_{k}^{(2)}+{i\over 2}-{i\pi\over 2\mu'})\big)\over \sh\big({\mu'\over 2}(\tilde{\lambda}+\lambda_{k}^{(2)}-{i\over 2}+{i\pi\over 2\mu'})\big)}{\ch\big({\mu'\over 2}(\tilde{\lambda}-\lambda_{k}^{(2)}+{i\over 2}+{i\pi\over 2\mu'})\big)\over \ch\big({\mu'\over 2}(\tilde{\lambda}-\lambda_{k}^{(2)}-{i\over 2}-{i\pi\over 2\mu'})\big)} \non \\
&\times& {\sh\big({\mu'\over 2}(\tilde{\lambda}+\lambda_{l}^{(1)}+{i\over 2}+{i\pi\over 2\mu'})\big)\over \sh\big({\mu'\over 2}(\tilde{\lambda}+\lambda_{l}^{(1)}-{i\over 2}-{i\pi\over 2\mu'})\big)}{\ch\big({\mu'\over 2}(\tilde{\lambda}-\lambda_{l}^{(1)}+{i\over 2}-{i\pi\over 2\mu'})\big)\over \ch\big({\mu'\over 2}(\tilde{\lambda}-\lambda_{l}^{(1)}-{i\over 2}+{i\pi\over 2\mu'})\big)}
\label{f}
\ee  
After evaluating the rest of the integrals, (\ref{amplitudes3}) becomes
\be
\alpha(\tilde{\lambda},a_{+})\alpha(\tilde{\lambda},a_{-}) &=& S_{0}(\tilde{\lambda})^{2}S_{1}(\tilde{\lambda},a_{-})S_{1}(\tilde{\lambda},a_{+})\prod_{k=1}^{{p-1\over 2}}\prod_{l=1}^{{p+1\over 2}}\sqrt{f(\lambda_{k}^{(2)},\lambda_{l}^{(1)},\tilde{\lambda})f(\lambda_{k}^{(2)},\lambda_{l}^{(1)},\tilde{\lambda}+{i\pi\over \mu'})} \non \\
\label{amplitudes4}
\ee
where 
\be
S_{0}(\tilde{\lambda}) &=& {1\over \pi}\ch(\mu'\tilde{\lambda})\prod_{n=0}^{\infty}{\Gamma\Big[{1\over \nu -1}(4n+1-2i\tilde{\lambda})\Big]\Gamma\Big[{1\over \nu -1}(4n+3-2i\tilde{\lambda})+1\Big]\over \Gamma\Big[{1\over \nu -1}(4n+1+2i\tilde{\lambda})\Big]\Gamma\Big[{1\over \nu -1}(4n+3+2i\tilde{\lambda})+1\Big]}\non \\
&\times& {\Gamma\Big[{1\over \nu -1}(4n+4+2i\tilde{\lambda})\Big]\Gamma\Big[{1\over \nu -1}(4n+2i\tilde{\lambda})+1\Big]\over \Gamma\Big[{1\over \nu -1}(4n+4-2i\tilde{\lambda})\Big]\Gamma\Big[{1\over \nu -1}(4n-2i\tilde{\lambda})+1\Big]}\non \\
&\times& {\Gamma^{2}\Big[{1\over \nu -1}(2n-i\tilde{\lambda})+{1\over 2}\Big]\Gamma^{2}\Big[{1\over \nu -1}(2n+1+i\tilde{\lambda})+{1\over 2}\Big]\over \Gamma^{2}\Big[{1\over \nu -1}(2n+2+i\tilde{\lambda})+{1\over 2}\Big]\Gamma^{2}\Big[{1\over \nu -1}(2n+1-i\tilde{\lambda})+{1\over 2}\Big]}
\label{S1}
\ee
\be
S_{1}(\tilde{\lambda},a_{\pm}) &=& {1\over \pi}\sqrt{\ch(\mu'(\tilde{\lambda}+{i\over 2}(\nu-2a_{\pm})))\ch(\mu'(\tilde{\lambda}-{i\over 2}(\nu-2a_{\pm})))}\non \\
&\times& \prod_{n=0}^{\infty}{\Gamma\Big[{1\over \nu -1}(2n+1+i\tilde{\lambda}-{1\over 2}(\nu-2a_{\pm}))+{1\over 2}\Big]\Gamma\Big[{1\over \nu -1}(2n+1+i\tilde{\lambda}+{1\over 2}(\nu-2a_{\pm}))+{1\over 2}\Big]\over \Gamma\Big[{1\over \nu -1}(2n+1-i\tilde{\lambda}-{1\over 2}(\nu-2a_{\pm}))+{1\over 2}\Big]\Gamma\Big[{1\over \nu -1}(2n+1-i\tilde{\lambda}+{1\over 2}(\nu-2a_{\pm}))+{1\over 2}\Big]}\non \\
&\times& {\Gamma\Big[{1\over \nu -1}(2n-i\tilde{\lambda}-{1\over 2}(\nu-2a_{\pm}))+{1\over 2}\Big]\Gamma\Big[{1\over \nu -1}(2n-i\tilde{\lambda}+{1\over 2}(\nu-2a_{\pm}))+{1\over 2}\Big]\over \Gamma\Big[{1\over \nu -1}(2n+2+i\tilde{\lambda}-{1\over 2}(\nu-2a_{\pm}))+{1\over 2}\Big]\Gamma\Big[{1\over \nu -1}(2n+2+i\tilde{\lambda}+{1\over 2}(\nu-2a_{\pm}))+{1\over 2}\Big]} 
\label{S2}
\ee
 
\noindent The values of the ``extra'' roots are dependent on the hole rapidity, $\tilde{\lambda}$ and the boundary parameters, $a_{\pm}$. Hence, it is sensible to expect a relation between these ``extra'' roots, $\Big\{\lambda_{k}^{(2)}, \lambda_{l}^{(1)}\Big\}$, the boundary parameters, $a_{\pm}$ and the hole rapidity, $\tilde{\lambda}$. Consequently, one needs to express the right hand side of (\ref{constraint}) in terms of purely $a_{\pm}$ and $\tilde{\lambda}$ to complete the derivation. To look for this additional relation, we begin with the information contained in the difference of the two densities, $\rho^{(1)}(\lambda) - \rho^{(2)}(\lambda)$. This leads to the following,
\be
\rho_{diff}(\lambda) &=& \rho^{(1)}(\lambda) - \rho^{(2)}(\lambda)\non \\
&=& {1\over N} R_{-}(\lambda)
\label{diffden}
\ee
where $R_{-}(\lambda)$ has the following Fourier transform,
\be
\hat R_{-}(\omega) &=& {1\over 1-\hat a_{2}(\omega)}
\Big[-\hat a_{1+2a_{-}}(\omega) - \hat a_{1-2a_{-}}(\omega) - \hat a_{1+2a_{+}}(\omega) - \hat a_{1-2a_{+}}(\omega) \non \\ 
&-&  2\hat b_{2}(\omega)\big(\sum_{k=1}^{{p-1\over 2}}\cos(\lambda_{k}^{(2)}\omega) - \sum_{l=1}^{{p+1\over 2}}\cos(\lambda_{l}^{(1)}\omega)\big)\Big]
\label{Rminus}
\ee
Analogous to (\ref{Yang})) one gets
\be
{1\over N}{d \h_{diff}(\lambda)\over d\lambda} = {1\over N} R_{-}(\lambda)
\label{Yang4} 
\ee
where $\h_{diff}(\lambda)=\h^{(1)}(\lambda)-\h^{(2)}(\lambda)$ and ${1\over N}{d \h_{diff}(\lambda)\over d\lambda} = \rho_{diff}(\lambda)$. 
Further, integrating (\ref{Yang4}) with respect to $\lambda$, taking limits of integration from $0$ to $\tilde{\lambda}$ as before, one finds 
\be
\h_{diff}(\tilde{\lambda}) &=& \h^{(1)}(\tilde{\lambda}) - \h^{(2)}(\tilde{\lambda})
= \int_{0}^{\tilde{\lambda}} d\lambda R_{-}(\lambda) 
\label{Yang5} 
\ee
Since $\h^{(1)}(\tilde{\lambda}) = \h^{(2)}(\tilde{\lambda})\in$ positive integer, using the fact that $R_{-}(\lambda)$ is an even function of $\lambda$ and exponentiating (\ref{Yang5}) we get 
\be
e^{\int_{-\tilde{\lambda}}^{\tilde{\lambda}} d\lambda R_{-}(\lambda)} = g(\tilde{\lambda},a_{+})g(\tilde{\lambda},a_{-})\prod_{k=1}^{{p-1\over 2}}\prod_{l=1}^{{p+1\over 2}}\sqrt{{f(\lambda_{k}^{(2)},\lambda_{l}^{(1)},\tilde{\lambda})\over f(\lambda_{k}^{(2)},\lambda_{l}^{(1)},\tilde{\lambda} + {i\pi\over \mu'})}}  = 1
\label{Rminus2}
\ee 
where $g(\tilde{\lambda},a_{\pm}) \equiv \sqrt{{\ch({i\mu'\over 2}(\nu-2a_{\pm})) + i\sh(\mu'\tilde{\lambda})\over \ch({i\mu'\over 2}(\nu-2a_{\pm})) - i\sh(\mu'\tilde{\lambda})}}$.
\noindent Next, an important observation is the following relation (as $N\rightarrow \infty$),  
\be
\prod_{k=1}^{{p-1\over 2}}\prod_{l=1}^{{p+1\over 2}}f(\lambda_{k}^{(2)},\lambda_{l}^{(1)},\tilde{\lambda}) = -1
\label{fconstraint}
\ee
for which we provide numerical support in Table 1. Although the results shown in Table 1 are computed for the case where the hole appears to the right of the largest ``sea'' root, we find similar results for other hole locations.         
From (\ref{Rminus2}) and (\ref{fconstraint}), it also follows that
\be
\prod_{k=1}^{{p-1\over 2}}\prod_{l=1}^{{p+1\over 2}}f(\lambda_{k}^{(2)},\lambda_{l}^{(1)},\tilde{\lambda}+{i\pi\over \mu'}) = -\bigg(g(\tilde{\lambda},a_{+})g(\tilde{\lambda},a_{-})\bigg)^{2}
\label{fconstraint2}
\ee

\begin{table}[htb] 
  \centering
  \begin{tabular}{|c|c|c|}\hline
    $N$ & $\prod_{k=1}^{{p-1\over 2}}\prod_{l=1}^{{p+1\over 2}}f(\lambda_{k}^{(2)},\lambda_{l}^{(1)},\tilde{\lambda})$, $p = 3$  & $\prod_{k=1}^{{p-1\over 2}}\prod_{l=1}^{{p+1\over 2}}f(\lambda_{k}^{(2)},\lambda_{l}^{(1)},\tilde{\lambda})$, $p = 5$\\
    \hline
    24      & -0.999364 + 0.0356655 i      & -0.999334 + 0.036496 i \\
    32      & -0.999421 + 0.0340133 i      & -0.999333 + 0.036522 i \\
    40      & -0.999466 + 0.0326686 i      & -0.999334 + 0.036486 i \\
    48      & -0.999502 + 0.0315413 i      & -0.999337 + 0.036419 i \\
    56      & -0.999532 + 0.0305749 i      & -0.999340 + 0.036336 i \\  
    64      & -0.999558 + 0.0297318 i      & -0.999343 + 0.036243 i \\  
    
    \hline
   \end{tabular}
   \caption{$\prod_{k=1}^{{p-1\over 2}}\prod_{l=1}^{{p+1\over 2}}f(\lambda_{k}^{(2)},\lambda_{l}^{(1)},\tilde{\lambda})$ for $p = 3$ ($a_{+}=2.1$, $a_{-}= 1.6$) 
    and $p = 5$ ($a_{+}=3.3$, $a_{-}= 2.7$), 
    from numerical solutions based on $N = 24$\,,$32$\,,\ldots\,,$64$.}
  \label{c2table}
\end{table}
\noindent We stress here that the values of $\lambda_{k}^{(2)}$ and $\lambda_{l}^{(1)}$ used in computations above strictly satisfy the Bethe equations (\ref{BAEII1}). Finally, we can rewrite (\ref{amplitudes4}) as
\be
\alpha(\tilde{\lambda},a_{+})\alpha(\tilde{\lambda},a_{-}) &=& S_{0}(\tilde{\lambda})^{2}S_{1}(\tilde{\lambda},a_{-})S_{1}(\tilde{\lambda},a_{+})g(\tilde{\lambda},a_{+})g(\tilde{\lambda},a_{-})
\label{amplitudes5}
\ee
up to a rapidity-independent phase factor. Subsequently, the complete expression for each boundary's scattering amplitude is given by (up to a rapidity-independent phase factor)
\be
\alpha(\tilde{\lambda},a_{\pm}) = S_{0}(\tilde{\lambda})S_{1}(\tilde{\lambda},a_{\pm})g(\tilde{\lambda},a_{\pm})
\label{eigenvalue1}
\ee
where $+$ and $-$ again denotes right and left boundaries respectively. 

\subsection{Eigenvalue for the one-hole state with 2-string}\label{w2s}

We now consider the one-hole state with a 2-string, reviewed in Section \ref{sec:2string}. The computation of the eigenvalue for this state is identical to the one given above. Hence, we skip the details and present the result. Analogous to (\ref{amplitudes3}), we have
\be
\beta(\tilde{\lambda},a_{+})\beta(\tilde{\lambda},a_{-}) &=& \exp\bigg\{2\int_{0}^{\infty}{d\omega\over \omega}\sh(2i\tilde{\lambda}\omega)\Big[{2\sh(3\omega/2)\sh((\nu-2)\omega/2)\over \sh(2\omega)\sh((\nu-1)\omega/2)}\non \\ 
&+& {\sh(\omega)\over \sh((\nu-1)\omega)\ch(\omega)} + {\sh((-\nu+2a_{-}-1)\omega)\over 2\sh((\nu-1)\omega)\ch(\omega)}\non \\ &+& {\sh((\nu-2a_{-}-1)\omega)\over 2\sh((\nu-1)\omega)\ch(\omega)} + (a_{-}\rightarrow a_{+})\non \\ 
&+& {\sh(\omega)\over \sh((\nu-1)\omega)}(\sum_{k=1}^{{p-3\over 2}}\cos(2\lambda_{k}^{(2)}\omega) + \sum_{l=1}^{{p-1\over 2}}\cos(2\lambda_{l}^{(1)}\omega))\non \\
&+& {\sh(2\omega)\over \sh((\nu-1)\omega)}(\ch(2i\lambda_{0}^{(1)}\omega) + \ch(2i\lambda_{0}^{(2)}\omega))\Big]\bigg\}
\label{amplitudes7}
\ee
which after evaluating the integrals yields 
\be
\beta(\tilde{\lambda},a_{+})\beta(\tilde{\lambda},a_{-}) &=& S_{0}(\tilde{\lambda})^{2}S_{1}(\tilde{\lambda},a_{-})S_{1}(\tilde{\lambda},a_{+})w(\lambda_{0}^{(1)},\tilde{\lambda})w(\lambda_{0}^{(2)},\tilde{\lambda})\non \\
&\times& \prod_{k=1}^{{p-3\over 2}}\prod_{l=1}^{{p-1\over 2}}\sqrt{f(\lambda_{k}^{(2)},\lambda_{l}^{(1)},\tilde{\lambda})f(\lambda_{k}^{(2)},\lambda_{l}^{(1)},\tilde{\lambda}+{i\pi\over \mu'})}
\label{amplitudes8}
\ee
where 
\be
w(\lambda_{0}^{(a)},\tilde{\lambda}) = \sqrt{{\ch(\mu'(\tilde{\lambda}+\lambda_{0}^{(a)}+i))\over \ch(\mu'(\tilde{\lambda}+\lambda_{0}^{(a)}-i))}{\ch(\mu'(\tilde{\lambda}-\lambda_{0}^{(a)}+i))\over \ch(\mu'(\tilde{\lambda}-\lambda_{0}^{(a)}-i))}}\,,\quad a = 1\,, 2\,,
\label{w}
\ee
As before, $(a_{-}\rightarrow a_{+})$ represents two additional terms which are the same as the third and fourth terms in the integrand of (\ref{amplitudes7}), but with $a_{-}$ replaced by $a_{+}$. We proceed to make the following conjecture (as $N\rightarrow \infty$) to complete the derivation.   
\be
w(\lambda_{0}^{(1)},\tilde{\lambda})w(\lambda_{0}^{(2)},\tilde{\lambda})\prod_{k=1}^{{p-3\over 2}}\prod_{l=1}^{{p-1\over 2}}\sqrt{f(\lambda_{k}^{(2)},\lambda_{l}^{(1)},\tilde{\lambda})f(\lambda_{k}^{(2)},\lambda_{l}^{(1)},\tilde{\lambda}+{i\pi\over \mu'})} &=& g(\tilde{\lambda}+{i\pi\over \mu'},a_{+})g(\tilde{\lambda}+{i\pi\over \mu'},a_{-})\non \\
\label{conjec}
\ee
Like (\ref{fconstraint}), we provide numerical support for (\ref{conjec}) in Table \ref{c4table} where we compute the ratio $\phi \equiv {d_{1}\over d_{2} }$, where $d_{1}$ and $d_{2}$ are the left hand side and the right hand side of (\ref{conjec}) respectively, for systems up to 64 sites. We believe this supports the validity of (\ref{conjec}) at $N \rightarrow \infty$. The values of $\lambda_{k}^{(2)}\,, \lambda_{l}^{(1)}\,, \lambda_{0}^{(1)}$ and $\lambda_{0}^{(2)}$ used in computations are obtained by solving numerically the Bethe equations (\ref{h1even2}) and (\ref{h2even2}) for the ``sea'' roots and  (\ref{BAEII1}) for the ``extra'' roots. The correctness and validity of such numerical solutions are checked by comparing them with the ones obtained from McCoy's method for smaller number of sites, e.g., $N = 2\,, 4$ and $6$ \footnote{We are only able to use McCoy's method to exactly solve for the Bethe roots for systems up to only 6 sites due to computer limitations.}. We stress here that although the results obtained in Table \ref{c4table} are computed for $\tilde{J} = 1$, namely the case where the hole appears close to the origin, similar results are found for other hole locations, e.g., $\tilde{J} = 2\,, 3\,,\ldots$.

\begin{table}[htb] 
  \centering
  \begin{tabular}{|c|c|c|}\hline
    $N$ & $\phi$, $p = 3$  & $\phi$, $p = 5$\\
    \hline
    24      &  0.967073 + 0.254500 i      &  0.990295 + 0.138982 i \\
    32      &  0.981063 + 0.193688 i      &  0.994434 + 0.105361 i \\
    40      &  0.987716 + 0.156259 i      &  0.996308 + 0.085849 i \\
    48      &  0.991392 + 0.130928 i      &  0.997674 + 0.068166 i \\
    56      &  0.993634 + 0.112654 i      &  0.998065 + 0.062174 i \\  
    64      &  0.995102 + 0.098852 i      &  0.998407 + 0.056428 i \\  
    
    \hline
   \end{tabular}
   \caption{$\phi$ for $p = 3$ ($a_{+}=2.1$, $a_{-}= 1.6$) 
    and $p = 5$ ($a_{+}=3.2$, $a_{-}= 2.7$), 
    from numerical solutions based on $N = 24$\,,$32$\,,\ldots\,,$64$.}
  \label{c4table}
\end{table}

Using (\ref{conjec}), the other eigenvalue for the Yang matrix (\ref{YangMatrix2}) becomes
\be
\beta(\tilde{\lambda},a_{+})\beta(\tilde{\lambda},a_{-}) = S_{0}(\tilde{\lambda})^{2}S_{1}(\tilde{\lambda},a_{+})S_{1}(\tilde{\lambda},a_{-})g(\tilde{\lambda}+{i\pi\over \mu'},a_{+})g(\tilde{\lambda}+{i\pi\over \mu'},a_{-})
\label{eigenvalue2}
\ee
hence giving the following for each boundary's scattering amplitude (up to a rapidity-independent phase factor),
\be
\beta(\tilde{\lambda},a_{\pm}) = S_{0}(\tilde{\lambda})S_{1}(\tilde{\lambda},a_{\pm})g(\tilde{\lambda}+{i\pi\over \mu'},a_{\pm})
\label{eigenvalue2b}
\ee
 
\subsection{Relation to boundary sine-Gordon model}\label{bsG}

Next, we briefly review Ghoshal-Zamolodchikov's results for the one boundary sine-Gordon theory \cite{GZ}. We borrow conventions used in \cite{ABNPT, CSS}. Ghoshal-Zamolodchikov's results imply that the right and left boundary $S$ matrices $R(\theta\,; \eta_{\pm}, \vartheta_{\pm},\gamma_{\pm})$ are given by 
\be
R(\theta\,; \eta, \vartheta, \gamma) = r_{0}(\theta)\ 
r_{1}(\theta\,; \eta, \vartheta)\ M(\theta\,; \eta, \vartheta, 
\gamma) \,,
\label{boundSmatrix}
\ee
where $M$ has matrix elements
\be
M(\theta\,; \eta, \vartheta, \gamma) =
\left( \begin{array}{cc}
m_{11} & m_{12} \\
m_{21} & m_{22}
	\end{array} \right) \,,
\ee
where $(\eta_{\pm}, \vartheta_{\pm}, \gamma_{\pm})$ are the Ghoshal-Zamolodchikov's IR parameters and $\theta$ is the hole-rapidity. Further,
\be
m_{11} &=& \cos \eta \cosh \vartheta \cosh (\tau \theta)
+ i \sin \eta \sinh \vartheta  \sinh (\tau \theta) \,, \non \\
m_{22} &=& \cos \eta \cosh \vartheta \cosh (\tau \theta)
- i \sin \eta \sinh \vartheta  \sinh (\tau \theta) \,, \non \\
m_{12} &=&i e^{i \gamma} \sinh(\tau \theta) 
\cosh (\tau \theta) \,, \non \\
m_{21} &=&i e^{-i \gamma} \sinh(\tau \theta) 
\cosh (\tau \theta) \,.
\label{matrixelements}
\ee 
where $\tau = {1\over \nu-1}$ is the bulk coupling constant. The scalar factors have the following integral representations \cite{ABNPT, CSS} 
\be
r_{0}(\theta) &=& \exp \left\{ 2i\int_{0}^{\infty} {d\omega\over \omega}
\sin (2\theta \omega/ \pi) {\sinh ((\nu-2)\omega/2) \sinh(3\omega/2)\over
\sinh((\nu-1)\omega/2) \sinh(2\omega)} \right\} \,, \non \\
r_{1}(\theta\,; \eta, \vartheta) &=& {1\over \cos \eta \cosh \vartheta}
\sigma(\eta, \theta)\ \sigma(i\vartheta, \theta) \,,
\label{r0r1}
\ee
where
\be
\sigma(x, \theta) = \exp \left\{ 2\int_{0}^{\infty} {d\omega\over \omega}
\sin((i\pi -\theta) \omega/(2\pi)) \sin(\theta \omega/(2\pi))
{\cosh ((\nu-1)\omega x/\pi) \over
\sinh((\nu-1)\omega/2) \cosh(\omega/2)} \right\} \,.
\ee
Our result (\ref{amplitudes5}) and (\ref{eigenvalue2}) agree with the eigenvalues of $R(\theta\,; \eta_{+}, \vartheta_{+},
\gamma_{+})R(\theta\,; \eta_{-}, \vartheta_{-},\gamma_{-})$, provided we make the following identification,
\be
\eta_{\pm} &=& {\mu'\over 2}(\nu-2a_{\pm}) \non \\
\theta &=& \pi\tilde{\lambda}
\label{IRandlattice}
\ee
In addition to (\ref{IRandlattice}), one should also take $\vartheta_{\pm} = \gamma_{\pm} = 0$, since they are related to the lattice parameters that appear in the spin chain Hamiltonian, (\ref{Hamiltonian}) which have been set to zero. Refer to the discussion following (\ref{bulkHamiltonian}). The same expression is given in \cite{ABNPT} for the corresponding open XXZ spin chain with nondiagonal boundary terms but with a constraint among the boundary parameters, hence suggesting that (\ref{IRandlattice}) holds true in general. 
As noted above, the eigenvalues, (\ref{eigenvalue1}) and (\ref{eigenvalue2b}) agree with the sine-Gordon boundary {\it S} matrix eigenvalues. Hence the two eigenvalues can be related as follows, 
\be
{\alpha(\tilde{\lambda},a_{\pm})\over \beta(\tilde{\lambda},a_{\pm})} = {\ch({i\mu'\over 2}(\nu-2a_{\pm})) + i\sh(\mu'\tilde{\lambda})\over \ch({i\mu'\over 2}(\nu-2a_{\pm})) - i\sh(\mu'\tilde{\lambda})}\label{ratio}
\ee       

\section{Discussion} 

Based on a recently proposed Bethe ansatz solution for an open spin-$1/2$ XXZ spin chain with nondiagonal boundary terms, we have derived the boundary scattering amplitude (equation (\ref{eigenvalue1})) for a certain one-hole state. We used a conjectured relation between the ``extra'' roots and the hole rapidity, namely (\ref{fconstraint}), which we verified numerically. This result agrees with the corresponding {\it S} matrix result for the one boundary sine-Gordon model derived by Ghoshal and Zamolodchikov \cite {GZ}, provided the lattice and IR parameters are related according to (\ref{IRandlattice}). We obtained the second eigenvalue (\ref{eigenvalue2b}) by considering an independent one-hole state with a 2-string. This scattering amplitude, derived for the one-hole state with 2-string also agrees with Ghoshal-Zamolodchikov's result following conjecture (\ref{conjec}), which we verified numerically and identification (\ref{IRandlattice}). It would be interesting to derive (\ref{fconstraint}) and (\ref{conjec}) analytically. 

It will also be interesting to study the excitations for the more general case of the open XXZ spin chain, namely with six arbitrary boundary parameters and arbitrary anisotropy parameter, and derive its corresponding {\it S} matrix. Solutions (spectrums) have been proposed for the general case, using the representation theory of q-Onsager algebra \cite{Koizumi} and the algebraic-functional method \cite{gall}. However, Bethe Ansatz solution for this general case has not been found so far although such a solution has been proposed lately for the XXZ spin chain with six boundary parameters at roots of unity \cite{MNS2}. In addition to the bulk excitations, one can equally well look at boundary excitations although this can be rather challenging even for the simpler case of spin chains with diagonal boundary terms. It is therefore our hope that some of these issues are addressed in future publications.   

\section*{Acknowledgments}

I would like to thank R.I. Nepomechie for his invaluable advice, suggestions 
and comments during the course of completing this work. I also fully appreciate the
financial support received from the Department of Physics, University of Miami. I also thank the referee for his/her helpful questions and suggestions that further helped to appropriately revise the paper.


\begin{thebibliography}{99}

\bibitem{Yang}
C. N. Yang,
``Some exact results for the many-body problem in one dimension with repulsive delta-function interaction,''
{\it Phys. Rev. Lett.} {\bf 19}, 1312 (1967)

\bibitem{ZZ}
A. B. Zamolochikov and Al. B. Zamolodchikov,
``Factorized S-matrices in two dimensions as the exact solutions of certain relativistic quantum field theory models,''
{\it Ann. Phys.} {\bf 120}, 253 (1979)

\bibitem{Baxter}
R. J. Baxter,
``Partition function of the eight-vertex lattice model,'' 
{\it Ann. Phys.} {\bf 70}, 193 (1972);  
{\ J. Stat. Phys.} {\bf 8}, 25 (1973);
``Exactly Solved Models in Statistical Mechanics,'' 
(Academic Press) (1982) 

\bibitem{Cherednik}
I. V. Cherednik, 
``Factorizing particles on a half-line and root systems,''
{\it Theor. Math. Phys.} {\bf 61}, 977 (1984)

\bibitem{GZ}
S. Ghoshal and A. B. Zamolodchikov, 
``Boundary S-matrix and boundary state in two-dimensional 
integrable quantum field theory,''
{\it Int. J. Mod. Phys.} {\bf A9}, 3841 (1994) 
[{\it Erratum ibid.} {\bf A9}, 4353 (1994)][{\tt hep-th/9306002}]

\bibitem{Korepin}
V. E. Korepin,
``Direct calculation of the S-matrix in the massive Thirring model,''
{\it Theor. Math. Phys.} {\bf 41}, 953 (1979)

\bibitem{A-D}
N. Andrei and C. Destri, 
``Dynamical symmetry breaking and fractionization in a new integrable model,''
{\it Nucl. Phys.} {\bf B231}, 445 (1984)

\bibitem{FT}
L. D. Fadeev and L. A. Takhtajan,
``Spectrum and scattering of excitations in the one-dimensional isotropic Heisenberg model,''
{\it J. Sov. Math.} {\bf 24}, 241 (1984)

\bibitem{S-F}
P. Fendley and H. Saleur,
``Deriving Boundary S-matrices,''
{\it Nucl. Phys.} {\bf B428}, 681 (1994)
[{\tt hep-th/9402045}]
    
\bibitem{GMN}
M. T. Grisaru, L. Mezincescu and R. I. Nepomechie,
``Direct calculation of the boundary S matrix for the open Heisenberg chain,''
{\it J. Phys.} {\bf A28}, 1027 (1995)
[{\tt hep-th/9407089}]

\bibitem{AN}
A. Doikou and R. I. Nepomechie,
``Direct calculation of breather S matrices,''
{\it J. Phys.} {\bf A32}, 3663 (1999)
[{\tt hep-th/9903066}]

\bibitem{LMSS}
A. LeClair, G. Mussardo, H. Saleur and S. Skorik,
``Boundary energy and bound states in integrable quantum field theories,''
{\it Nucl. Phys.} {\bf B453}, 581 (1995)
[{\tt hep-th/9503227}]

\bibitem{Anep}
C. Ahn and R.I. Nepomechie,
``Finite-size effects in the XXZ and sine-Gordon models with two boundaries,''
{\it Nucl. Phys.} {\bf B676}, 637 (2004)
[{\tt hep-th/0309261}]

\bibitem{AhnBella}
C. Ahn, M. Bellacosa and F. Ravanini,
``Excited states NLIE for sine-Gordon model in a strip with Dirichlet boundary conditions,''
{\it Phys. Lett.} {\bf B595}, 537 (2004)
[{\tt hep-th/0312176}]

\bibitem{ABNPT}
C. Ahn, Z. Bajnok, R.I.Nepomechie, L. Palla and G. Takacs,
``NLIE for hole excited states in the sine-Gordon model with two boundaries,''
{\it Nucl. Phys.} {\bf B714}, 307 (2005)
[{\tt hep-th/0501047}]

\bibitem{doikou}
A. Doikou,
``Generic boundary scattering in the open XXZ chain,''
[{\tt 0711.0716}]

\bibitem{KB}
A. Klumper and M. T. Batchelor,
``An analytic treatment of finite-size corrections in the spin-$1$ antiferromagnetic XXZ chain,''
{\it J. Phys.} {\bf A23}, L189 (1990); 
A. Klumper, M. T. Batchelor and P. A. Pearce,
``Central charges of the 6- and 19-vertex models with twisted boundary conditions,''
{\it J. Phys.} {\bf A24}, 3111 (1991)

\bibitem{Destri}
C. Destri and H. J. De Vega,
``New thermodynamic Bethe ansatz equations without strings,''
{\it Phys. Rev. Lett.} {\bf 69}, 2313 (1992);
C. Destri and H. J. De Vega,
``Unified approach to thermodynamic Bethe-ansatz and finite-size corrections for lattice models and field-theories,''
{\it Nucl. Phys.} {\bf B438}, 413 (1995)
[{hep-th/9407117}];

\bibitem{Fio}
D. Fioravanti, A. Mariottini, E. Quattrini and F. Ravanini,
``Excited state Destri - De Vega equation for sine-Gordon and restricted sine-Gordon models,''
{\it Phys. Lett.} {\bf B390}, 243 (1997)
[{hep-th/9608091}]

\bibitem{DD}
C. Destri and H. J. De Vega,
``Non-linear integral equation and excited-states scaling functions in the sine-Gordon model,''
{\it Nucl. Phys.} {\bf B504}, 621 (1997);
[{hep-th/9701107}]

\bibitem{FRT}
G. Feverati, F. Ravanini and G. Takacs,
``Nonlinear Integral Equation and Finite Volume Spectrum of Sine-Gordon Theory,''
{\it Nucl. Phys.} {\bf B540}, 543 (1999);
[{hep-th/9805117}];
G. Feverati, F. Ravanini and G. Takacs,
``Truncated Conformal Space at c=1, Nonlinear Integral Equation and Quantization Rules for Multi-Soliton States,''
{\it Phys. Lett.} {\bf B430}, 264 (1998);
[{hep-th/9803104}]

\bibitem{Feverati}
G. Feverati,
`` Finite Volume Spectrum of Sine-Gordon Model and its Restrictions,''
[{hep-th/0001172}]

\bibitem{MN}
R. Murgan and R.I. Nepomechie, 
``Generalized $T-Q$ relations and the open XXZ chain,'' 
{\it J. Stat. Mech.} {\bf P08002} (2005)
[{\tt hep-th/0507139}]

\bibitem{MNS}
R. Murgan, R.I. Nepomechie and C. Shi,
``Boundary energy of the open XXZ chain from new exact solutions,'' 
{\it Annales Henri Poincare} {\bf 7}, 1429 (2006)
[{\tt hep-th/0512058}]

\bibitem{Murgan}
R. Murgan,
``Finite-size correction and bulk hole-excitations for special case of an open XXZ chain with nondiagonal boundary terms at roots of unity,''
{\it JHEP} {\bf 05}, 069 (2007) 
[{\tt 0704.2265}]

\bibitem{dVega}
H.J. de Vega and A. Gonz\'alez-Ruiz, 
``Boundary K-matrices for the
six vertex and the $n(2n-1)$ $A_{n-1}$ vertex models,'' 
{\it J. Phys.} {\bf A26}, L519 (1993) 
[{\tt hep-th/9211114}]

\bibitem{CSS}
J. -S. Caux, H. Saleur and F. Siano,
``The two-boundary sine-Gordon model,'' 
{\it Nucl. Phys.} {\bf B672}, 411 (2003)
[{\tt cond-mat/0306328}]

\bibitem{Koizumi}
P. Baseilhac and K. Koizumi, 
``Exact spectrum of the XXZ open spin chain from the q-Onsager algebra representation theory,''
{\it J. Stat. Mech.} {\bf P09006} (2007)
[{\tt hep-th/0703106}]

\bibitem{gall}
W. Galleas,
``Functional relations from the Yang-Baxter algebra: Eigenvalues of the XXZ model with non-diagonal twisted and open boundary conditions,''
{\it Nucl. Phys.} {\bf B790}, 524 (2008)
[{\tt 0708.0009}]

\bibitem{MNS2}
R. Murgan, R.I. Nepomechie and C. Shi
``Exact solution of the open XXZ chain with general integrable
boundary terms at roots of unity,'' 
{\it J. Stat. Mech.} {\bf P08006} (2006)
[{\tt hep-th/0605223}]

\end{thebibliography}
\end{document}